# Synthesis of water-soluble melanin

Erika Soares Bronze-Uhle,[1,2,*] Marina Piacenti-Silva,[1] João Vitor Paulin,[2] Chiara Battocchio[3] and Carlos Frederico de Oliveira Graeff[1]

[1] Department of Physics, FC-UNESP, Av. Eng. Luiz Edmundo Carrijo Coube 14-01, 17033-360 Bauru, Brazil

[2] UNESP - Universidade Estadual Paulista, POSMAT - Programa de Pós-Graduação em Ciência e Tecnologia de Materiais, Av. Eng. Luiz Edmundo Carrijo Coube 14-01, Bauru, SP, Brazil

[3] Department of Sciences, University "Roma Tre", Via della Vasca Navale 79, 00146 Rome, Italy

\* Corresponding Author: Tel: +55-14-3103-6000, r. 6375; Fax: +55-14-31036094.
*E-mail address*: eriuhle@fc.unesp.br

**Abstract**

Melanins are promising materials for organic bioelectronics devices like transistors, sensors and batteries. However, in general, melanin either natural or synthetic has low solubility in most solvents. In this study, the chemical structural changes of melanin synthesized, by the auto oxidation of L-DOPA, are analyzed for a new synthetic procedure using a reactor with oxygen pressure of 4 atm. UV-Vis spectroscopy, FTIR, C-NMR, XPS and TEM are use to characterize the material. Under oxygen pressure, the synthesis of melanin is accelerated and the polymer obtained is found to have higher carbonyl groups compared to conventional synthetic melanin. As a consequence it has higher homogeneity and is soluble in water. To explain these findings a reaction mechanism is proposed based on current melanogenesis models.

# 1. Introduction

Melanin is a unique natural product, starting for its properties and abundance, it can be found from primitive life organisms like fungus to human beings. [1] The most common form of melanin is eumelanin, an irregular heteropolymer obtained by the copolymerization of derivatives of 5,6-dihydroxy-indole-2-carboxylic acid (DHICA) and 5,6-dihydroxy-indole (DHI) and its various redox forms states, see Figure 1. [2,3] They present unique physical-chemical properties such as broad absorbance throughout the UV-visible region, strong non-radiative relaxation of photoexcited electronic states, hybrid ionic-electronic conduction, intrinsic free radical character, ion storage and biocompatibility which inspires its use in bioelectronics or biomedical applications. [1]

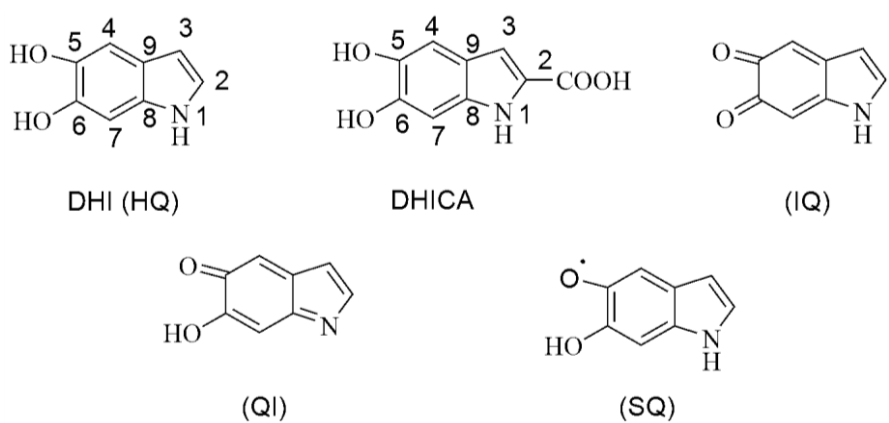

**Fig. 1**. The basic building blocks of melanin

Though recognized as a potential electronic material since at least the 70's, [4] the quest for technological applications has a major barrier: production of good adherent films. At present, many research groups are finding different solutions to this problem. Thin film with good properties was produced using simple techniques such as spin coating and drop casting by solubilizing the pigment in aqueous solution with ammonia. [5] However, alkali treatment can degrade melanin chemical structure. [6] Thus alternatives have been developed, for example, the work of Guin et al. [7] presents a simple method to overcome these difficulties. Starting with the conventional and easy to prepare synthetic melanin, they have shown that thin films with good adherence and controllable thickness can be made using LbL. It has also been reported that it is possible to obtain melanin films from electrochemical deposition methods, [8] electrospray, [9] matrix-assisted pulsed laser evaporation (MAPLE) [10] and the direct polymerization of L-DOPA, [11], DHI, [12] and dopamine [13] on the substrate.

An alternative to obtain good quality films is to modify eumelanin structure preserving the pigment main properties but increasing its solubility. The advantage of this approach is that all processing techniques used up to now for organic semiconducting polymers device production would be available for melanin. Not just that, many routine advance techniques dedicated to thin film or solution characterization would also be available. In the last 15 years soluble melanin derivatives were synthesized using different functionalized groups: sulfone, [14,15] benzyl and octyl ester, [16] glycation [17] and triethyleneglycol. [18]

In this study, we propose an alternative synthetic approach for a soluble melanin without any exotic functionalization using molecular oxygen as the oxidant. Molecular oxygen is considered an ideal oxidant it has low cost, is abundant and environmentally friendly, having thus academic and industrial interest, especially in the context of "green chemistry". It is found that in addition to solubility the synthesis is faster.

**2. Materials and methods**

All the commercially available chemicals were purchased from Acros or Sigma-Aldrich and were used without further purification. In what concerns reproducibility, more than 5 different syntheses were performed and similar optical and structural properties were obtained. The synthesis of melanin was carried out using well-known procedures briefly described.[15] Initially 1.0 g of L-3,4-dihydroxyphenylalanine (L-DOPA) was dissolved in 200 mL of deionized water (Milli-Q), in the presence of ammonium hydroxide ($NH_4OH$) until the pH of the mixture was between 8 and 10. This mixture was stirred for 3 days at room temperature bubbling air using an air pump. The process of melanin polymerization was followed using UV–Vis spectroscopy.[19] Upon synthesis completion, the solution was poured in a dialysis membrane of 5000 D and submerged in Mili-Q water, in order to remove residues and then dried at 80 °C. The product obtained is a black powder and will be referred as *Mel*. For the synthesis of melanin under oxygen pressure a stainless steel reactor was used with a capacity of 150 mL. 0.30 g of L-DOPA and 400 μL of ammonium hydroxide were mixed in 60 mL of Milli-Q water, (pH = 10). After sealing the reactor, oxygen was introduced until the internal pressure reached 4 atm. The reaction was kept under stirring and the process of melanin polymerization was followed using UV–Vis spectroscopy.[19] After 15 hours, the synthesis was complete and the solution was purified in the same way described in the previous section. The product obtained is referred as *Mel-P*. We have tested the synthesis without ammonium hydroxide. The reaction proceeds very slowly and after 22 days no significant amount of material had been formed, demonstrating the importance of $NH_4OH$ in the synthesis process.

To verify the synthesis progression, periodically an aliquot was withdrawn from the solution for UV-Vis Spectroscopy analysis using a Shimadzu UV-Vis-Spectrophotometer (UVmini-1240). Aliquots removed during the synthesis of Mel and Mel-P was diluted at a ratio of 0.07/1 in $H_2O$. Absorption spectra from 290 to 800 nm were recorded at regular time intervals using a quartz cuvette of 1 cm path length, until the reaction ended. The criteria for defining the reaction end were discussed elsewhere.[19] FTIR spectroscopy measurements were made on a Bruker Vertex 70 between 4000 and 400 cm$^{-1}$, at room temperature in Attenuated Total Reflectance (ATR) mode. The band at 2360 cm$^{-1}$ is coming from atmospheric $CO_2$, due to incomplete purging with $N_2$. $^{13}$C CP/MAS analyses were performed on a Bruker Avance III 400 MHz spectrometer equipped with a 4mm CP/MAS probe, operating at 100.5 MHz for $^{13}$C. The $^{13}$C CP/MAS spectra of the solid melanins were obtained by means of the cross-polarization technique (Cross-Polarization Magic Angle Spinning - CPMAS) with contact time of 2 ms, repetition time of 5 s and MAS rotation frequency of 5kHz. Two Pulse Phase Modulation (tppm) proton decoupling was used. XPS analysis was performed in a self built instrument, consisting of a UHV preparation and analysis chamber, equipped with a 150 mm mean radius hemispherical electron analyzer with a four-element lens system with 16-channel detector giving a total instrumental resolution of 1.0 eV as measured at the Ag $3d_{5/2}$ core level. MgKα non-monochromatic X-ray radiation (hν = 1253.6eV) was used for acquiring core level spectra of all samples (C1s, N1s, S2p and O1s). The spectra were energy referenced to the C1s signal of aliphatic C atoms having a binding energy (BE) of 285.00 eV. Atomic ratios were calculated from peak intensities by using Scofield's cross-section values and calculated λ factors. Curve-fitting analysis of the C1s, N1s, S2p and O1s spectra was performed using Gaussian profiles as fitting functions, after subtraction of a Shirley-type background. For TEM images, the sample was prepared by dropping a suspension of melanin in isopropyl alcohol on a copper grid (300 mesh) coated with carbon film and dried in air. Image acquisition was done using a transmission electron microscope FEI Tecnai G2F20 operating at 200 kV.

**3. Results**

The processes of melanin polymerization were followed using UV–Vis spectroscopy. Mel and Mel-P synthesis are characterized by a gradual increase in absorption throughout the spectrum. It was observed that the reaction end is reached earlier in Mel-P (15 hours) when compared to Mel (3 days).

Figure 2 shows the absorption spectra of Mel and Mel-P. The absorbance of both melanins is characterized by a featureless broadband spectrum that increases exponentially toward the ultraviolet. [2] This broad absorption spectrum is associated to a sum of individual absorptions of different oligomeric units, suggesting large chemical disorder. [2] Although Mel and Mel-P show almost the same behavior, Mel-P has an absorbance smaller than Mel especially above 350 nm. The insert in Figure 2 shows a picture of Mel-P and Mel dispersed in Mili-Q water. As can be seen, Mel-P is soluble, while Mel not. We have tested the solubility in different solvents, namely toluene, isopropanol, tetrahydrofuran, chloroform, ehtyl acetate, acetone, methanol, acetonitrile, dimethylformamide, N-methil-2-pyrrolidone, dimethyl sulfoxide and MiliQ water (pH = 7.0). In all tests 5.0 mg of melanin powder was diluted in 1.0 mL, Mel-P is found to be soluble only in water. We also have used the procedure used by Cicco et al [18], in which the solubility of a melanin derivative is tested by filtration using a nylon membrane (0.45 µm). A sample is considered soluble when the filter does not retain the passage of the particles. In our case, a filter with a regenerated cellulose membrane (0.2 µm) and a solution of 5.0 mg/mL in MiliQ water was used. Again, Mel-P proved to be soluble in water in this test.

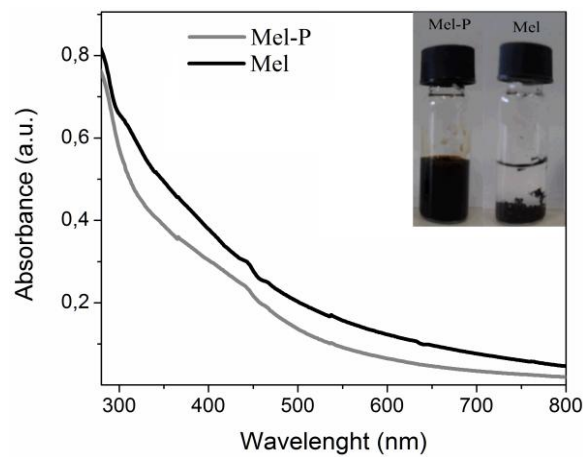

**Fig. 2**. Absorbance spectra obtained from Mel and Mel-P powder suspension in water. In the insert a photo showing the solubility of the two samples.

Figure 3 shows the FTIR spectra for Mel and Mel-P. These spectra were taken after the synthesis was considered complete. The spectra are similar for both materials, however Mel-P has some prominent peaks. As already reported in the literature, the broad band between 2500 and 3500 cm$^{-1}$ is associated to –OH bond stretching of the DHI and DHICA derivatives. [20] The bands at 1600 and 1430 cm$^{-1}$ are related to phenol C=C and stretching of C–O from ionized carboxylic acid. Between 1180-1280 cm$^{-1}$ can be seen the stretching bands of C–OH from phenolic or carboxyl OH. The carbonyl group of DHICA (-COOH) or 5,6-indole quinones or semiquinones is evident from the stretching at approximately 1724-1486 cm$^{-1}$. The results shown in Figure 3 indicate that Mel-P has an increased carbonyl content compared to Mel, by the evident increase of the bands at 1396 and 1186 cm$^{-1}$, referent to the stretching modes of ionized carbonyl acid and phenolic or carboxyl –OH groups. [20]

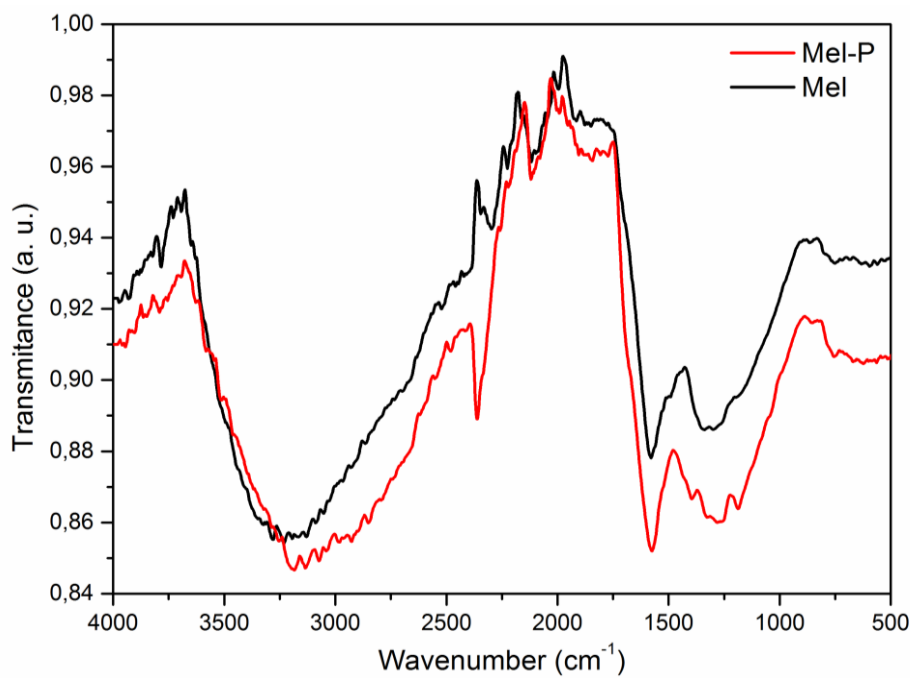

**Fig. 3**. FTIR spectra of Mel-P and Mel**.**

Figures 4(a) and 4(b) shows $^{13}$C CP/MAS-NMR spectra for Mel and Mel-P samples. The spectra show broad resonances due the heterogeneity in the polymer creating a dispersion of chemical shifts. [21,22] Thus, the large number of overlaps does not allow the resolution of individual bands of carbon in melanin. By convention, the signals in the $^{13}$C CP/MAS NMR of Mel and Mel-P can be divided into three main resonance regions due to different carbon melanin oligomers: (1) aliphatic groups (0-90 ppm) from unreacted L-DOPA; (2) aromatic carbons (95-155 ppm) including pyrrole and indole carbons (CH$_x$, C-O, C$_2$NH); and (3) carbonyl groups (160-200 ppm) from carboxylic acids and carboxyl of indolequinones (COO$^-$ and C=O). Furthermore, protonated (95-114 ppm) and non-protonated (110-147 ppm) carbons can be observed. [23,24] The spectra of $^{13}$C NMR shown in Figures 4(a) and 4(b) are similar in shape, however the area of the carbons signals have differences depending on synthesis indicating structural changes in polymeric material, especially in the region regarding the carbonyls groups.

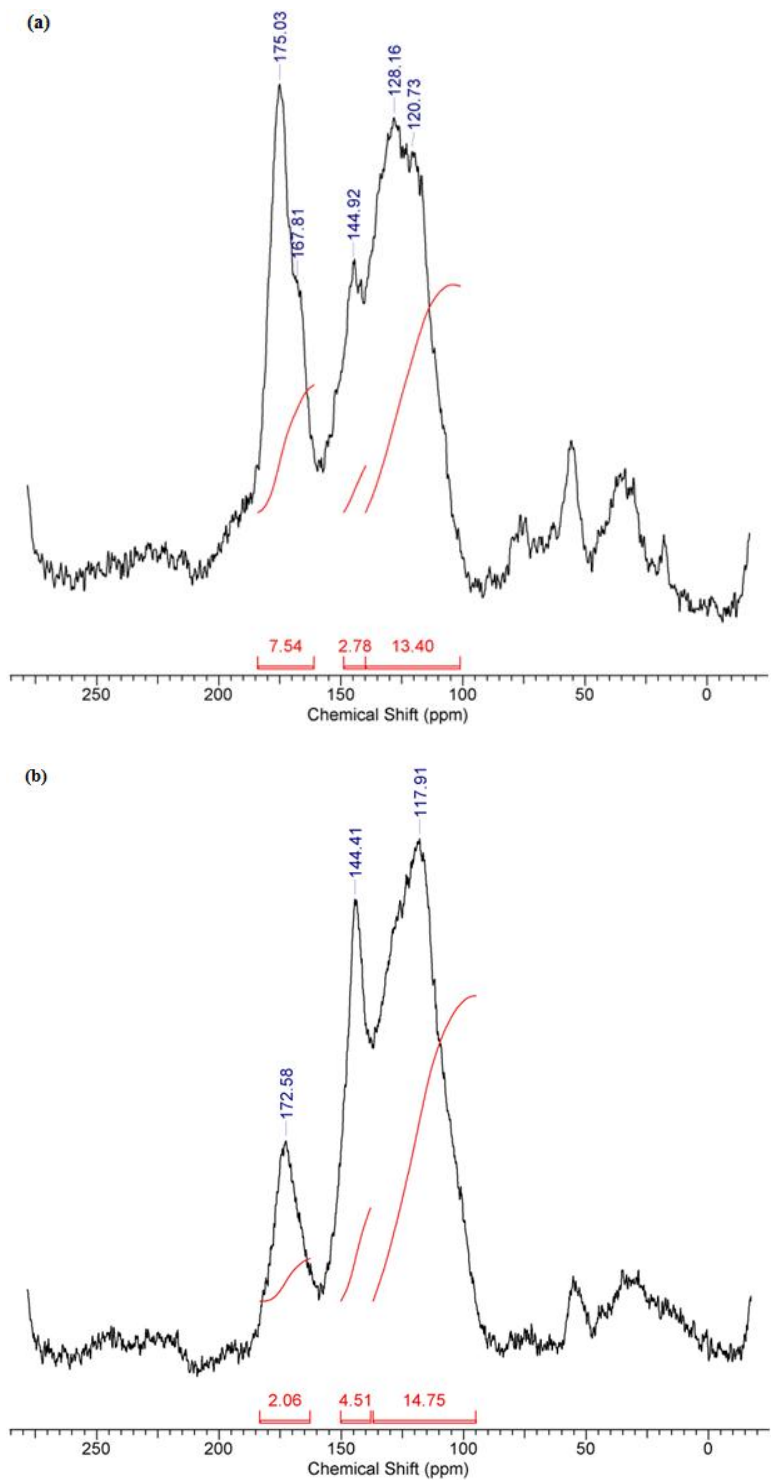

**Fig. 4.** $^{13}$C CP/MAS NMR spectra of Mel-P (a) and Mel (b) showing the individual signals of melanin.

XPS signals of Mel and Mel-P are presented in Figure 5. A fitting of the C1$s$ core level spectra was performed. The results are in good agreement with the literature. [9]

Five contributions to C1s spectra are observed: the peak at lower binding energies, centered in 285.0 eV due to C-C; the component at about 286.3 eV is assigned to C-N; the peak at 287.5 eV to C-O; the two peaks at higher binding energies are attributed to –COOH and COO⁻ arising from DOPA. The same features are observed in both samples, but with different relative intensities. The intensity of the different contributions to the C1s photoemission was calculated as described in Abbas (2009). [9] The results are presented in Table 1.

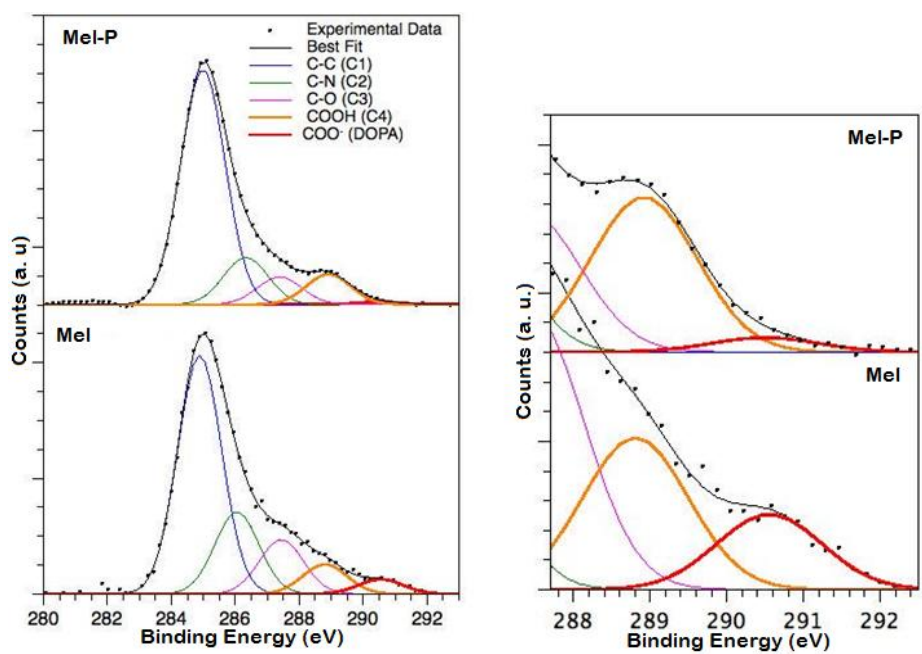

**Fig. 5.** XPS signals of Mel-P and Mel from the C1*s* core levels. The fittings are presented as solid lines, while the experimental data with solid circles.

**Table 1**: C1s Photoemission intensity values (normalized) as calculated from the fitting procedure shown in Fig. 5.

| Sample | signal | BE (eV) | Photoemission C1s Intensity (normalized) | Assignment |
|---|---|---|---|---|
| Mel-P | C1s | 285.00 | 6.0 | C-C |
|  |  | 286.34 | 1.1 | C-N |
|  |  | 287.39 | 0.9 | C-O |
|  |  | 288.92 | 0.78 | COOH |
|  |  | 290.50 | 0.07 | COO⁻ (DOPA) |

| | | 285.00 | 5.0 | C-C |
|---|---|---|---|---|
| | | 286.17 | 1.6 | C-N |
| Mel | C1s | 287.56 | 1.4 | C-O |
| | | 288.92 | 0.53 | COOH |
| | | 290.67 | 0.27 | COO$^-$ (DOPA) |

Figure 6 shows TEM images of Mel and Mel-P. It can be seen that the structures of all melanins are similar, regardless of the synthetic route, with grains of different sizes and very irregular. The magnitude of the grains observed in the images is in agreement with the literature, which describes melanin partially composed of nanodiscs pilled up having approximately 2 nm in diameter and 1 nm in height with a tendency to bond forming clusters of different dimensions. [3] From this figure Mel-P has a more homogeneous structure or in other words it has smaller clusters.

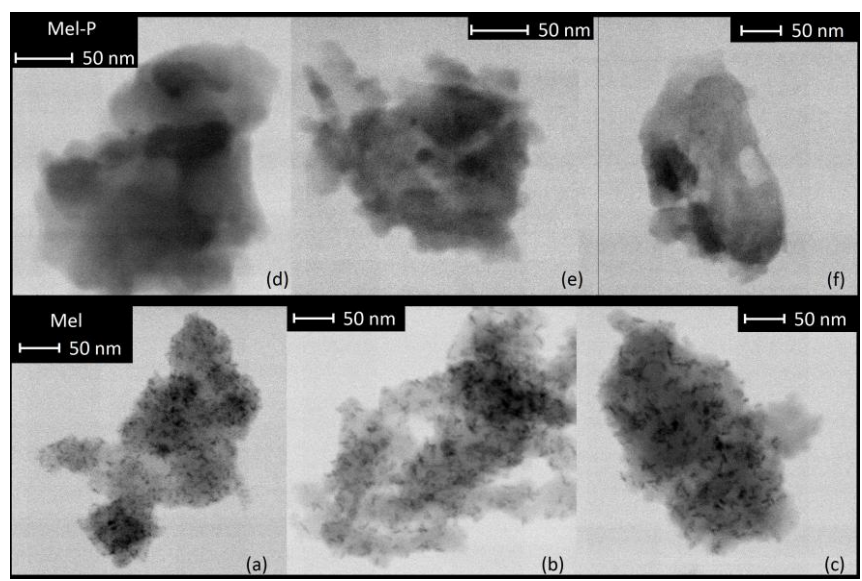

**Fig. 6.** Transmission Electron Microscopy Images of Mel-P and Mel.

### 4. Discussion

Due to the insolubility and the difficulty in extracting melanin from its natural sources, different *in vitro* routes were proposed. [11,12,14,16,18,25,26] Melanin can be synthesized from the oxidation of tyrosine or L-3,4-dihydroxyphenylalanine (L-DOPA) via enzymatic reactions or auto-oxidation of L-DOPA in alkaline aqueous solution, by bubbling air. [6] These melanins syntheses occur in an uncontrolled manner, thereby culminating in a polymeric material with a high degree of chemical heterogeneity with different proportions of carboxylated units (DHICA and derivatives) and decarboxylated monomers (DHI and derivatives). Different structures for melanin macromolecule were proposed, but details of how these units connect are still under debate, but details of how these units connect are still under debate. However, it is believed that melanin present a small planar graphene like structures, made of DHI, DHICA and oxidized forms, stacked into two or three sheets with inter-planar distance of 3.7–4.0 Å. [3].

*In vivo*, melanins are synthesized from the auto oxidation of tyrosine through the Raper-Mason mechanism. [27,28] Natural melanin also presents a combination of DHI and DHICA in varying proportions depending on the source of melanin. In general, Sepia melanin contains DHICA/DHI ratios higher than 50%. On the other hand, synthetic melanins generally have a higher proportion of DHI with respect to DHICA, and the ratio varies greatly depending on synthesis conditions. [29] The differences in DHI and DHICA proportion both in natural and synthetic melanin affect the physicochemical and biological function of the material. Carboxyl group of DHICA is strongly connected to the anti-chelating and oxidant capacity of this pigment, in addition the presence of COOH or DHICA/DHI ratio changes the polymerization process, limiting the size of the polymer. [1] Therefore, the conditions under which the synthesis occur are extremely important, since small changes in the reaction condition can affect the thermodynamics, altering the polymer structure and consequently the properties of the final product. Since the biological functions of melanins are closely associated to their chemical composition and structure, it is essential to understand better the synthesis and its final products, since the relation structure/properties in this class of materials are still poorly understood.

The specific factors involved in the production of DHICA or DHI and other functional groups in different proportions are still unknown, however, the oxidation of organic compounds induced by air (autoxidation) are well understood, and play an important role in the generation of free radicals, particularly in biological media, through reactive oxygen species (ROS). [30] Reactive metabolites such as ROS produced *in vivo*, can oxidize various biomolecules and organic compounds such as lipids, proteins, nucleic acids, thiols, phenols, sulfides, etc. With the oxygen constantly present in our environment, autoxidation processes can chemically alter biological substrates and cellular events associated, so the knowledge of the mechanisms of ROS formation and reaction, such as in quinones (L-DOPA) are required in order to understand the structural changes in the products and kinetics involved. [31]

Much of the current knowledge about the structure and function of melanin comes from studies with synthetic material and is therefore fundamental to obtain and characterize melanins using different synthetic procedures, since for example there is a marked structural difference between natural melanin and its various synthetic equivalents.

In the melanogenesis basically two enzymes derived from tyrosine, named tyrosinase related protein-1 (Tyrp1) and tyrosinase related protein-2 (Tyrp2), are active in the oxidation process. The initial step involves the enzymatic O-hydroxylation of tyrosinase to 3,4-dihydroxyphenylalanine (DOPA) and/or oxidation of L-DOPA to dopaquinone, catalyzed by tyrosinase. Dopaquinone is a very reactive intermediate that spontaneously undergoes rearrangement and cyclization generating dopachrome [1].

Until the 80's, it was believe that the subsequent reactions followed the Raper-Mason mechanism, in which spontaneous decarboxylation occurred generating DHI and a number of other unstable compounds.[32] However, this mechanism did not explain the high concentration of carbonyl compounds such as DHICA in natural melanin. To explain the existence of DHICA it was proposed that in the enzymatic oxidation (*in vivo*) Tyrp2, also called dopachrome tautomerase, catalyzes the tautomerization of dopachrome into DHICA, and Tyrp1 catalyzes the oxidation of DHICA promoting its polymerization (see Scheme 1).[33] Thus, Tyrp2 explains why natural melanins possess DHICA/DHI greater than 50%.

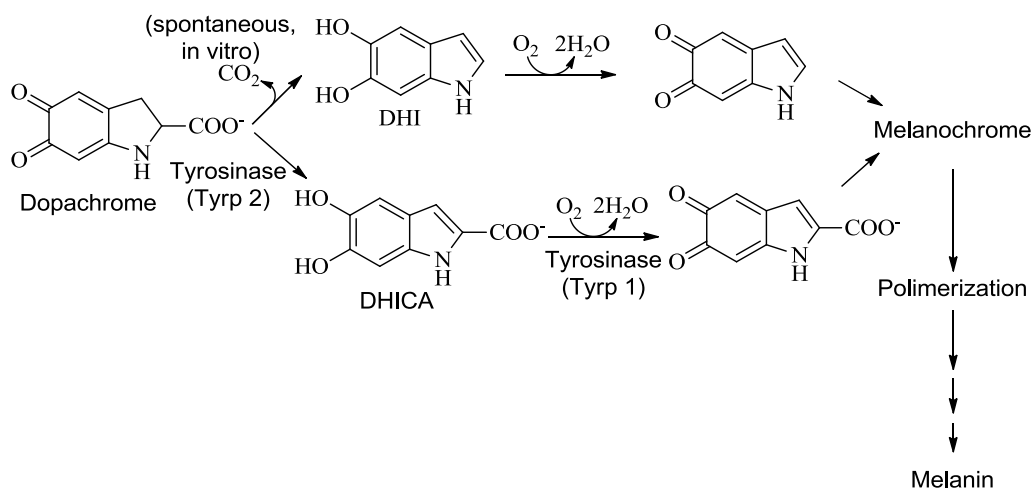

**Scheme 1**. Action of tyrosinase enzyme on dopachrome.

During melanin synthesis, both *in vitro* and *in vivo*, a number of steps in the o-quinones oxidation process (enzymatically catalyzed or not) produce hydrogen peroxide ($H_2O_2$), which is accumulated in the reaction medium.[34] The $H_2O_2$ formation mechanisms and its function during melanogenesis is not fully understood, but studies indicate that it suffers direct action of tyrosinase.

Some kinetic studies indicate that tyrosinase may also react with hydrogen peroxide through two processes: Catalase (i.e. conversion of $H_2O_2$ to ½$O_2$ and $H_2O$) and Peroxygenase ($H_2O_2$-dependent oxygenation of substrates). Both activities can be explained by the existence of oxy-tyrosinase peroxo dicopper (II) species, which act as a common reactive intermediate.

Tyrosinase is a copper monooxygenase enzyme present in plants, fungi, bacteria and animals, which act catalyzing the oxygenation of the substrate by the action of oxygen. During the oxidation process, the oxidized forms of this enzyme (oxy-forms) involve peroxo dicopper (II) species responsible for phenol oxygenation and catechol dehydrogenation reactions producing hydroxobridged dicopper(II) species (met-form) in the reaction medium. Met-tyrosinase reacts with catechols yielding o-quinones and deoxy-tyrosinase (dicopper(I)form). The deoxy-tyrosinase reacts with $O_2$, generating oxy tyrosinase completing the catalytic cycle (see Scheme 2). The o-quinone products are spontaneously converted to melanin pigment and $H_2O_2$. [34–36]

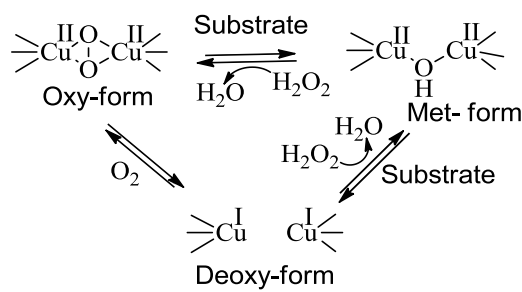

**Scheme 2.** Tyrosinase derived species during the oxidation reactions.

In catalase, these species are formed by the reaction of met-tyrosinase with $H_2O_2$ formed in the reaction medium. The oxy-tyrosinase complex, (peroxo)dicopper(II), reacts with $O_2$ generating deoxy-tyrosinase (dicopper(I)form), which then reacts with another $H_2O_2$ molecule generating again met-tyrosinase and $H_2O$, completing the catalytic cycle. When a phenolic substrate is added, the oxidized species of tyrozine, (peroxo)dicopper(II), attacks the substrate to induce phenolic C-O bond formation by oxygen transfer in a electrophilic aromatic substitution mechanism. [34–36]

In peroxygenase, the enzyme may catalyze the oxygenation of p-substituted phenols by $H_2O_2$. [33,37] Furthermore, studies realized by Wood et al, [37] show that low concentrations of $H_2O_2$ tend to activate the enzyme. Thus, many steps of the synthesis process are responsible for changing the relative concentration of DHI and DHICA in the final product.

Polymerization starts as soon as DHI and DHICA are formed. The relative concentration of DHI or DHICA influences strongly the process and consequently the optical and structural properties of melanin. [19,38] DHI rich melanin produces more heterogeneous structure, due to the various positions available in this monomer for polymerization. DHI can polymerize by the positions 2, 3, 4 and 7 (see Fig. 1) generating larger and branched macromolecules, while DHICA polymerizes only at positions 4 and 7, resulting in smaller and linear polymers. [39]

The differences in optical absorption seen in Figure 2 reflect the different oligomeric structures present in Mel-P and Mel due to different content of DHI and DHICA. [19,38] Oligomers with molecular weights of approximately 1000 amu exhibit less absorption compared with higher molecular weight oligomers, which indicates that Mel has higher molecular weight than Mel-P. Besides, melanins have a tendency to aggregate reducing the solubility and increasing the absorption and dispersion of light, which could also explain partially the higher absorbance of Mel compared to Mel-P. In Mel one expects an increase in the number of intermonomer connections, leading to more aggregation. [19,40] Therefore, considering the absorbance spectra obtained in Figure 2, it is believed that the Mel-P is composed of smaller structures, with a lower degree of aggregation, which also explains their better solubility, shown in insert in Figure 2.

Images of transmission electron microscopy, Figure 6, confirm the differences in aggregation between Mel and Mel-P. Mel has several regions that suggest quite large agglomerates, different from Mel-P, which is more homogeneous or at least composed of smaller aggregates. These features reinforce the hypothesis that in Mel-P prevails DHICA over DHI reducing aggregation and originating more regular structures.

The bands observed in FTIR spectra from Figure 3 are consistent with the literature, confirming that melanin was obtained in both procedures. [20] Even though melanin presents great structural complexity, small differences between the synthesized compounds were observed. The degree of conjugation of C=C, C=N and C=O or the size of the polymer, can affect the bond length and consequently the position of the bands. In Figure 3 a small red shift in the position of the spectrum bands is observed for Mel compared to Mel-P, showing a higher conjugation for this material. Stronger bands are observed in Mel-P compared to Mel in 1689 $cm^{-1}$, 1587 $cm^{-1}$ and 1396 $cm^{-1}$ compatible with a higher content of DHICA in this material.

$^{13}$C NMR and the XPS measurements corroborate the DHICA/DHI ratio increase. From $^{13}$C NMR spectra, Figure 4, Mel-P when compared to Mel, presents an area decrease in aromatic carbons (90-160 ppm) from 14.55 to 13.40 and in C-O aromatic carbons (144 ppm) from 4.51 to 2.78, suggesting the formation of more phenolic C=O, by a factor of 2. However, the principal modification is the large increase (a factor of 3) in carbons related to C=O and COO-, from 2.6 in Mel to 7.54 in Mel-P. As suggested by Yan Liu et al, [39] the relative increase in the amount of carbonyl and carboxylate groups versus aromatic carbons can be linked to an increase in the proportions of DHICA or to the oxidation of phenolic carbons. From the analysis it is understood that the large increase in carbonyl groups in Mel-P is related to the higher number of DHICA units in the final polymer. Our results suggest also that the material obtained under oxygen

pressure has a lower degree of polymerization (aromatic carbons decrease), in good agreement with UV-Vis absorption spectroscopy. [21,22] XPS also confirms this hypothesis. From Figure 5 and Table 1, the amount of -COOH and -COO⁻ groups per monomer can be estimated to be about 53% DHICA and 27% DOPA for Mel; while about 78% DHICA and 7% DOPA for Mel-P.

The main structural differences between Mel and Mel-P are primarily due to higher availability of oxygen in the latter. Thus, it is assumed that the larger oxygen partial pressure in the solution favored the production of a polymer with higher DHICA content, resembling the biological synthesis of melanin. It is known that in the biological environment the higher proportion of DHICA is associated with tyrosinase (Tyrp2, Scheme 1). In the synthetic process described herein, the oxidation of L-DOPA in the aqueous phase requires an alkaline medium (supplied by $NH_4OH$), suggesting an effect of hydroxyl ions in the oxidation reaction, in addition to the inherent capacity of oxidation of L-DOPA. In fact, the reaction does not proceed without $NH_4OH$, as described in section 2. With the increased oxygen pressure (4 atm) a higher oxidation activity is expected from the mass transfer of $O_2$ from the vapor phase to the aqueous phase. The alkaline medium is necessary to deprotonate primary alcohols, leading to cyclization and formation of Dopachrome or Leucodopachrome. The mechanism follows then a series of oxidative steps that lead to the formation of DHI and DHICA. In this process, the formation of $H_2O_2$ is expected, as well as in the biosynthetic process, as mentioned early. Remember that in the absence of the tyrosinase Tyrp2 higher proportions of DHI are obtained in the reaction medium [33], by the fast spontaneous decarboxylation.

In the biological reaction, the tyrosinase enzyme tends to consume $H_2O_2$ through catalysis, decreasing the concentration of peroxide in the reaction medium as a response to the oxidative stress.[41] Furthermore, the presence of $H_2O_2$ may be responsible for the cleavage of C-C bonds generating $CO_2$.[42,43] Thus we propose that in Mel synthesis, $H_2O_2$ induces C-C cleavage of the -COOH function forming $CO_2$ and thus more DHI over DHICA. Assuming that the synthesis of Mel-P is similar to Mel, in Mel-P the rate of L-DOPA oxidation increases as well as the decomposition of $H_2O_2$, originating $O_2$ and $H_2O$, which promotes even further the oxidation by increasing the concentration of dissolved oxygen in the reaction medium.[42] Consequently, the peroxide decomposition in the oxygen rich solution changes the products of the auto-oxidation reaction, a smaller number of -COOH will be cleaved, with lower $CO_2$ production and hence a greater production of DHICA relatively to DHI. Thus, the synthesis under oxygen pressure is associated with an increase in reaction velocity and selectivity toward DHICA.

The increase in carboxyl and carboxylates groups lead to noteworthy characteristics, DHICA rich pigments have higher chelating ability and oxidant capacity [1,39] and, as can be seen in Fig. 2, increases the solubility of melanin. The higher solubility of Mel-P is a pre-requisite to the formation of good quality thin films, as previously observed in DMSO synthesized melanin [14,44–46] and our own preliminary results (not shown here). On the other hand, carboxylic acid is already widely used as an anchoring group in organic electronics and some results suggested that it could improve the protonic conductivity of natural materials.[47] Finally, the smaller and apparently more ordered structure found in Mel-P, can be useful for bioelectronics applications, such as in biosensors.[48]

## 5. Conclusions

In this study, melanin was synthesized through the oxidative polymerization of L-DOPA under oxygen pressure of 4 atm, producing a material more soluble in water. A reaction mechanism was proposed based on the higher oxidative nature of the reaction medium including oxygen-induced decomposition of $H_2O_2$. It was observed that the oxygen pressure leads to an increase in the reaction velocity, reducing the time required for the synthesis by a factor of 5 compared to conventional synthesis performed at ambient pressure and air bubbling. In addition, melanin synthesized under oxygen pressure presents a higher content of carboxylated functions or DHICA/DHI ratio, making it similar to natural melanin, with an increase in water solubility. The solubility in water makes this synthesis path very attractive for thin film applications such as in biotransistors. The material obtained has also a more homogeneous structure as observed by TEM.

**Acknowledgments**

We would like to thank the Brazilian agencies CAPES, FAPESP, and CNPq for financial support, under contracts: FAPESP 2013/07296-2 and 2008/57872-1; CNPq 573636/2008-7. The authors would also thank Professor Alviclér Magalhães from Universidade Estadual de Campinas (UNICAMP) for $^{13}$C CP/MAS NMR analysis and Professor Paulo Marcos Donate from Universidade de São Paulo (USP) for the reactor system.